\documentclass[epj]{svjour}

\usepackage{graphicx}

\renewcommand{\d}{\mathop{\rm d}}

\begin{document}

\title{Symmetry breaking and restoring under high pressure:\\
the amazing behaviour of the ``simple'' alkali metals}

\author{G. G. N. Angilella \and F. Siringo \and R. Pucci}

\titlerunning{Friedel oscillations in alkalis at high pressure}

\authorrunning{G. G. N. Angilella \emph{et al.}}

\institute{Dipartimento di Fisica e Astronomia, Universit\`a di
   Catania, and Istituto Nazionale per la Fisica della Materia, UdR
   di Catania,\\ Via S. Sofia, 64, I-95123 Catania, Italy}

\date{\today}

\abstract{%
We argue that an ionic lattice surrounded by a Fermi liquid changes phase
   several times under pressure, oscillating between the symmetric
   phase and a low-symmetry dimerized structure, as a consequence of
   Friedel oscillations in the pair potential. 
Phase oscillations explain the tendency towards dimerization which has
   been recently reported for the light alkali metals under high
   pressure.
Moreover, a restoring of the symmetric phase is predicted for such
   elements at an even higher density.
\PACS{%
{71.20.Dg}{Alkali and alkaline earth metals}
\and
{62.50.+p}{High-pressure and shock-wave effects in solids and liquids}
\and
{64.70.Kb}{Solid-solid transitions}}}

\maketitle

\section{Introduction}

Some years ago we predicted \cite{Siringo:97} that under high pressure 
   the light alkali metals should be unstable
   towards less symmetric phases.
We also argued that the lowering of symmetry could give rise to a
   metal-insulator transition. 
We did not attempt to predict the structure of the high pressure
   phase, and unfortunately any such theoretical prediction is quite
   difficult, because of the small energy differences which separate
   most phases.
Our result followed the early suggestion that electron correlation in
   the alkalis could give rise to broken symmetry of the charge
   density wave (CDW) type, albeit incommensurate with the lattice
   \cite{Overhauser:85}.
The role of core repulsion in stabilizing the high pressure phases of
   the alkali had been also emphasized by McMahan
   \emph{et al.} \cite{Moriarty:82,McMahan:84}.

The prediction of a symmetry lowering was quite unexpected, as it is
   usually believed that an increase of pressure should give rise to
   an increase in the symmetry of the system.
Actually, under very high pressure the most likely state of matter is
   a uniform plasma, so that any system is expected to raise its
   symmetry at a sufficiently high pressure. 
However, many ``simple'' metals are still far from such a limiting
   behaviour, and their path towards the ultimate metallization may
   contain oscillations of their degree of symmetry.

More recently, both theoretical calculations \cite{Neaton:99} and 
   experimental findings \cite{Hanfland:99,Hanfland:00,Fortov:99} have 
   confirmed the tendency of the light alkali metals to lower their
   symmetry under high pressure.
Low symmetry phases have been observed for Li \cite{Hanfland:00} and
   Na \cite{Hanfland:private} (see also Ref.~\cite{Neaton:01}), whereas
   the eventual occurrence of a 
   metal-insulator transition has not been observed yet \cite{Fortov:99}. 
According to recent calculations \cite{Christensen:01}, both Li and Na 
   show a tendency towards the formation of atomic pairs, and a
   dimerized $oC$8 structure would be the most stable phase above
   165~GPa for Li, and above 220~GPa for Na.

The physical reason for dimerization is not evident. 
\emph{Ab initio} electronic structure calculations \cite{Rousseau:00}
   indicated a tendency towards ``distance alternation'', due to a
   sizeable overlap of $p\pi$ orbitals in the interstitial regions.
On the other hand, it has been noticed \cite{Christensen:01} that the increase of $s$-$p$
   hybridization could give rise to a low coordination number.  
However, even the fully dimerized phase is far from any standard
   covalent solid: the electron density is uniformly spread and almost
   constant, while the first and second neighbours distances are
   comparable. 
Such a dimerized phase is better described as a charge density wave in a high
   density metal rather than a molecular solid.
Now, at zero pressure both Li and Na are already well described by a
   simple degenerate Fermi liquid where the kinetic energy is the most
   relevant energy term \cite{AM}. 
At high density all the interactions should become smaller and smaller
   compared to the kinetic energy, and if all the energy terms are
   assumed to be monotonic functions of density, then nothing relevant 
   is expected to happen. 
On the other hand, the high density limit is where the Fermi liquid
   model should work better, thus it remains to be explained why a
   Fermi liquid should be unstable towards a charge density wave.

In this paper we address such a problem and show that an ionic lattice,
   surrounded by a Fermi liquid, oscillates between a simple metal and a
   charge density wave, several times with increasing density. 
We start in Sec.~\ref{sec:structural} by reviewing the evaluation of
   the structural energy of a solid lattice within linear response theory.
In Sec.~\ref{sec:alkali}, we later discuss the case of lithium and
   other alkali metals under high pressure, where the presence of
   Friedel oscillations in the screened pair interactions can justify
   the occurrence of an instability towards a broken-symmetry phase.
We eventually summarize in Sec.~\ref{sec:conclusions}.

\section{Structural energy of solids close to a phase instability}
\label{sec:structural}

The basic idea is foreshadowed in the seminal works of Pettifor
   \emph{et al.}
   \cite{Pettifor:70,Pettifor:84}, who discussed the possible
   existence of phase oscillations. 
The physical motivation is related to Friedel oscillations
   \cite{Friedel:54}, which characterize the screened interactions and
   give rise to a non-monotonic oscillating behaviour of the
   structural energy terms. 
A dimerization of the lattice can be regarded as a broken-symmetry
   phase described by an order parameter $y$ which vanishes in the
   symmetric phase. 
In most cases, $y$ can be taken to be the difference between first and
   second neighbours distances, and $y=0$ yields the monoatomic
   lattice.
The total structural energy of the system must be an even function
of $y$ and can be expanded as $U(y) \approx U(0) + \frac{1}{2}
   U^{\prime\prime} (0) y^2$.
Its main contribution comes from screened pair interactions which
   oscillate with a wavevector $q=2k_{\rm F}$, where $k_{\rm F}$ is
   the Fermi wavevector (Friedel oscillations). 
Thus the sign of the second derivative $U^{\prime\prime} (0)$ is
   dominated by the sign of the second derivative of the pair
   interaction with respect to the pair distance.
But an oscillating pair interaction yields oscillating derivatives,
   so that $U^{\prime\prime} (0)$ is expected to change sign as the
   pair distance decreases.
A positive sign $U^{\prime\prime} (0)>0$ corresponds to having a
   minimum for $y=0$ (symmetric phase); a 
   negative sign $U^{\prime\prime} (0)<0$ corresponds to having a
   relative maximum for $y=0$, thus indicating that 
   the symmetric phase is unstable towards a dimerized phase.
Moreover, we expect that with increasing density the sign of
   $U^{\prime\prime} (0)$ could change 
   several times, thus giving rise to phase oscillations between a
   symmetric and a dimerized lattice.

In order to make the above idea more quantitative, let us specialize
   to the light alkali metals.
We start by expressing the total energy per atom of a metal lattice in
   real-space formulation as \cite{Hafner:87}:
\begin{equation}
U = U_0 + \frac{1}{2} \sum_{i\neq j} \Phi (R_{ij} ),
\label{eq:totenergy}
\end{equation}
where the pair potential $\Phi (R_{ij} )$ measures the interaction
   between two ions located at sites $i$ and $j$ in the lattice,
   $R_{ij}$ being their mutual distance, and $U_0$ summarizes all
   the contributions independent of the lattice structure.
Within second order local pseudopotential theory, the pair potential
   can be written as \cite{Pettifor:84,Hafner:86}: 
\begin{equation}
\Phi (R) = \frac{Z^2}{R} \left( 1 + \frac{2}{\pi} \int_0^\infty h(q)
   v^2 (q) \frac{\sin qR}{q} \d q \right),
\label{eq:pairpot}
\end{equation}
where $Z$ is the atomic number,
\begin{equation}
h(q) = - \frac{\kappa^2}{q^2} \frac{\chi (q)}{\epsilon (q)} 
= - \kappa^2 \frac{\chi (q)}{q^2 + \kappa^2 [1-G(q)]\chi (q)} ,
\end{equation}
with $\epsilon(q)$ the dielectric screening function, $\kappa^2 = 4
   k_{\rm F} / \pi$ is the Thomas-Fermi screening parameter, $\chi(q)$
   is the normalized Lindhard susceptibility [$\chi(0)=1$], and $G(q)$ 
   takes into account for exchange and correlation corrections (local
   field corrections) to the electron-electron interaction
   \cite{Ichimaru:81}.
In the following, we shall take $Z=1$ for simplicity, which is a good approximation
   for the light alkali at ambient conditions.
In Eq.~(\ref{eq:pairpot}) we assume the `empty core' model for the
   ionic pseudopotential \cite{Ashcroft:66},
\begin{equation}
v(q) = \cos q R_c ,
\label{eq:pseudopot}
\end{equation}
where $R_c$ defines the radius of the atomic core.
The core radius $R_c$ is usually obtained by fitting
   Eq.~\ref{eq:pseudopot} against the value of the band gap $2|v({\bf
   g})|$.
We have checked that Eq.~(\ref{eq:pseudopot}) is also consistent with
   the calculated band structure of lithium both in the bcc
   \cite{Ching:74} and in the fcc structure \cite{Boettger:85}.
However, different sources of experimental data result in slightly
   different values of $R_c$ for a given element \cite{Cohen:70}.
Moreover, the value of $R_c$ can be effectively altered by chemical
   substitution (e.g. in alloys).
Therefore, in the following we will regard $R_c$ as a
   parameter ranging within given bounds for each element of
   interest.
The other independent parameter of the model is the electron
   spacing $r_s$, defined as $4\pi r_s^3 / 3 = N/V$, where $N/V$ is
   the conduction electron density.
Such parameter enters Eq.~(\ref{eq:pairpot}) through the Fermi
   momentum $k_{\rm F}$ in the Thomas-Fermi parameter $\kappa^2$, and
   can be used as a measure of pressure, with $r_s$ decreasing as
   pressure increases.
Moreover, for a given lattice structure, all inter-site distances
   $R_{ij}$ in Eq.~(\ref{eq:totenergy}) scale with $r_s$.

\begin{figure}
\centering
\includegraphics[height=0.9\columnwidth,angle=-90]{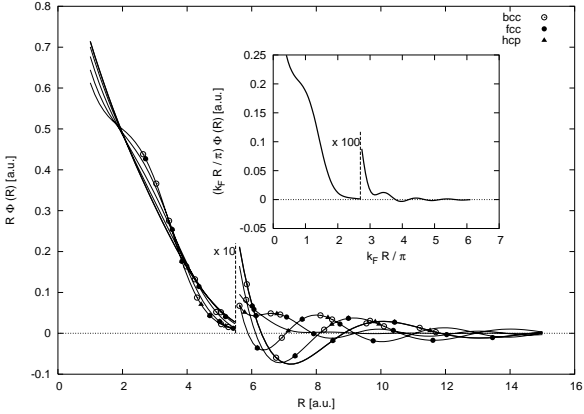}
\caption{%
Scaled pair potential $R\Phi(R)$ versus interatomic distance $R$.
\underline{\sl Main plot:} $R_c = 1.76$~a.u., $r_s = 1.5 -
   3.326$~a.u., with thicker line corresponding to $r_s = 3.326$~a.u.
Symbols correspond to nearest and farther neighbours locations in the
   bcc (open circles), fcc (full circles), and hcp (triangles)
   lattices, respectively.
\underline{\sl Inset:} Same as main plot for $R_c = 1.76$~a.u., $r_s =
   1.5$~a.u., but now with distances scaled by $k_{\rm F} /\pi$.
Notice that minima in $R\Phi(R)$ occur close to integer values of
   $k_{\rm F} R/\pi$ (Friedel oscillations).
}
\label{fig:pairpot}
\end{figure}

As a result of the singular behaviour of $h(q)$ at $q=2k_{\rm F}$,
   which in turn arises from the logarithmic discontinuity in the
   derivative of the Lindhard function $\chi(q)$, the Fourier
   transform in Eq.~(\ref{eq:pairpot}) endows the pair potential
   $\Phi(R)$ with an oscillating behaviour, with a characteristic
   length $\sim \pi/k_{\rm F}$ (Friedel oscillations, see
   Fig.~\ref{fig:pairpot}, inset). 
The analytical properties of the pair potential $\Phi(R)$ have
   been analyzed further in Ref.~\cite{Hafner:86}.
The actual value of the total energy $U$ in Eq.~(\ref{eq:totenergy})
   then depends on whether the distances of nearest and farther neighbours in
   the lattice lay close to maxima or minima in the plot of $\Phi(R)$
   (Fig.~\ref{fig:pairpot}).
For a fixed value of the pseudopotential parameter $R_c$, such
   distances can be actually shifted to lower values by decreasing
   $r_s$, i.e. by means of an applied pressure.
On the basis of such considerations, the stability of
   the crystal structures of several elements under pressure has been
   discussed in the past \cite{Hafner:83}.

\section{The case of lithium and other light alkali metals under high
   pressure}
\label{sec:alkali}

\begin{figure}
\centering
\includegraphics[bb=84 120 443 619,clip,width=0.7\columnwidth]{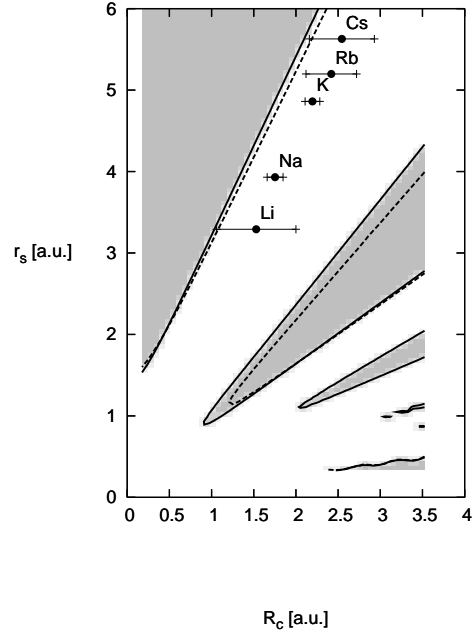}
\caption{%
Shaded regions in the plane of parameters $(R_c , r_s )$ are
   characterized by the condition $U_{yy} [x,0] < 0$, which
   signals an instability towards dimerization.
The regions corresponding to instability towards dimerization in the
   undistorted bcc phase ($x=0$) and in the $cI$16 phase ($x=0.05$) are
   bounded by continuous and broken lines, respectively.
Data for $(R_c , r_s )$ pairs for the light alkali metals at ambient
   pressure have been taken from
   Refs.~\protect\cite{Hafner:83,Cohen:70}.
Different methods of fitting pseudopotentials to the experimental data 
   result in a rather wide range for $R_c$, which is here displayed as
   an error bar.
}
\label{fig:phasediag}
\end{figure}

The phase diagram of the alkali at low pressure has been extensively
   addressed both experimentally and theoretically, with lithium
   crystallizing in the 9R phase at zero temperature and pressure
   \cite{Boettger:85,Lin:86,Holzapfel:96,Liu:99}.
At high pressure ($\sim 39$~GPa), a new structural phase (Pearson
   symbol $cI$16) has been recently detected \cite{Hanfland:00}.
Hanfland \emph{et al.} \cite{Hanfland:private} have recently observed
   bcc~$\to$~fcc~$\to$~$cI$16 transitions in Na. 
Thus we can regard the $cI$16 phase as our starting point for the
   following discussion on alkali metals.
Such a high-pressure phase is characterized by a bcc primitive cell,
   with an 8-atom basis \cite{Christensen:01}, and can be thought of
   as a distorted bcc phase, with a distortion parameter $x=0.045 -
   0.060$ \cite{Hanfland:00}.
The $cI$16 phase formally reduces to the bcc structure (`supercell' with 
   eight usual cubic cells) for $x=0$ \cite{Christensen:01}.
In the undistorted bcc phase, the lattice is composed of two
   interpenetrating cubic sublattices, $A$ and $B$, say.
Such classification applies to the $cI$16 structure as well, after
   distortion from the parent bcc lattice, but now with a basis of
   four atoms for each sublattice.
The $cI$16 phase by itself is not dimerized, and it has been recently
   predicted to be even superconducting \cite{Christensen:01a,Shimizu:02a}. 
We would like to
   test its stability with respect to a dimerized phase obtained
   by rigidly shifting the two sublattices each other 
   of a tiny amount $y$ along the $(111)$ direction. 
Here, $y$
   plays the role of the order parameter discussed above.
For a fixed value of the distortion parameter $x$, one can then think
   of expanding the total energy per atom Eq.~(\ref{eq:totenergy}) in
   powers of our `dimerization' parameter $y$,
\begin{equation}
U[x,y] = U_0 + U_{y} [x,0] y + \frac{1}{2} U_{yy} [x,0] y^2 +
   O(y^3 ),
\end{equation}
where $U_y = \partial U /\partial y$ etc.
Due to crystal symmetry, it can be proved analytically that $U_{y}
   [0,0] = 0$ exactly, whereas we numerically checked that $|U_y [x,0]
   /U_{yy} [x,0] | \ll 1$, for $x\ll 1$ \cite{note1}.
Therefore, an indication of instability towards `dimerization' is
   provided by the condition $U_{yy} [x,0] < 0$.
From Eqs.~(\ref{eq:totenergy}) and (\ref{eq:pairpot}), $U_{yy} [x,0]$
   can be expressed as
\begin{eqnarray}
U_{yy} [x,0] &=& \frac{1}{4} \sum_{{\bf n}\mu\nu}
\left[
\left( \Phi^{\prime\prime} (R^0_{{\bf n}\mu\nu} ) - \frac{1}{R^0_{{\bf
   n}\mu\nu}} \Phi^\prime (R^0_{{\bf n}\mu\nu} ) \right) 
\right.\times \nonumber\\
&&
\left.
\left( \frac{\partial R^0_{{\bf n}\mu\nu}}{\partial y} \right)^2 +
   \frac{3a^2}{R^0_{{\bf n}\mu\nu}} \Phi^\prime (R^0_{{\bf n}\mu\nu} )
\right],
\label{eq:derivata}
\end{eqnarray}
where $\vec{n}$ labels lattice points in the primitive bcc lattice,
   with lattice parameter $a$, $\mu$ and $\nu$ label the basis
   vectors in sublattice $A$ and $B$, respectively
   \cite{Christensen:01}, and $R^0_{{\bf 
   n}\mu\nu}$ denotes the mutual distances between lattice points for
   $y=0$.
Eq.~(\ref{eq:derivata}) has been evaluated on a finite lattice, large
   enough to reach full convergence.
Figure~\ref{fig:phasediag} displays our numerical results for $U_{yy}
   [x,0]$ as a function of parameters $(R_c , r_s )$, for $x=0$
   (undistorted bcc phase) and $x=0.05$ (representative value of the
   high-pressure $cI$16 phase
   according to Refs.\cite{Hanfland:00,Christensen:01}).

In the plane of parameters $(R_c, r_s)$ we find regions of instability
   towards dimerization where $U_{yy}[x,0]<0$
   (Fig.~\ref{fig:phasediag}, shaded areas). 
It is remarkable that at ambient pressure all the alkali metals are
   predicted to be stable in the symmetric phase.
As the core radius $R_c$ is related to the band gap, one would in
   general expect $R_c$ to be a (monotonic) function of the density
   parameter $r_s$.
More generally, a more quantitative analysis would require a non-local 
   pseudopotential, with a density-dependent range in reciprocal
   space.
However, within linear response
   theory, the long wavelength behaviour of the pseudopotential is
   what actually matters.
Indeed, the first zero of Eq.~\ref{eq:pseudopot} occurs at $q=\pi/2
   R_c$, which is very close to the smallest reciprocal lattice vector 
   of most metals.
Therefore, at the level of approximation implied by the present
   calculation, we can safely neglect the density dependence of $R_c$, 
   without qualitatively alter our main conclusions.
In this way, the effect of applied pressure on the phase point
   corresponding to each alkali metal in Fig.~\ref{fig:phasediag} can
   be tracked as a vertical shift at constant $R_c$, with increasing
   density.
However, the error bars for $R_c$ under normal conditions grossly provide
   order of magnitude boundaries for the generally expected variation
   of $R_c$ with density.

A sign change in $U_{yy}$ for a given alkali metal then corresponds to 
   a critical density.
The uncertainty on $R_c$ lends a wide window for the critical
   densities of Li and Na: the largest values (still compatible with
   the core radius data of Refs.~\cite{Hafner:83,Cohen:70}) are $r_s
   \sim 2 - 2.2$~a.u., which are not too far from the values observed by
   Hanfland \emph{et al.} \cite{Hanfland:00} for the onset of a dimerized
   phase for Li ($r_s\approx 2.1$~a.u. and $P=165$~GPa) and predicted by
   Christensen and Novikov \cite{Christensen:01} for the onset of a
   dimerized phase for Na ($r_s\approx 2.3$~a.u. and $P=220$~GPa).
However, we caution that such a critical density occurs for $r_s \sim
   R_c$, \emph{i.e.} close to the limits of applicability of the
   empty-core pseudopotential approximation, thus suggesting that more
   refined pseudopotentials should be employed in more realistic
   calculations.
The phase diagram of Fig.~\ref{fig:phasediag} also suggests that, with
   a further increase of density, both Li and Na should drop below the
   instability area, and a second transition would eventually restore
   the symmetry of the lattice.
This oscillating behaviour, from a symmetric phase to a less symmetric
   one and back to a restored symmetric phase, seems to be a general
   effect due to the presence of Friedel oscillations in the
   interatomic distance dependence of the pair potential.
Such oscillations in turn are a consequence of the existence of a
   Fermi surface at $k=k_{\rm F}$. 
Thus the tendency towards the formation of atomic pairs would be a
   signature of the Fermi liquid behaviour of the alkali metals.

Our method relies on linear response theory, which contains the
   Thomas-Fermi approximation as a long wavelength limit
   \cite{note:Hafner}. 
Both such methods are known to be reliable if the density gradients
   are not too strong. 
In large atoms, the Thomas-Fermi method yields results which only
   differ by 8\% from Hartree-Fock calculations, in space regions
   where the density gradient ${\nabla \rho/\rho}\approx 2.5/{a_0}$
   ($a_0$ is the Bohr radius) \cite{note:Slater}. 
Here, the typical density gradients are ${\nabla \rho/\rho}<
   0.5/{a_0}$, as reported by Ref.~\cite{Christensen:01} for the high
   pressure phase of Li at 165~GPa.
Thus, we expect linear response to be reliable in the high pressure 
   range, where eventual second order corrections are very small (for
   these gradients even the simple Thomas-Fermi approximation would
   deviate less than 1\% from Hartree-Fock calculations).

\section{Conclusions}
\label{sec:conclusions}

In summary, we have shown that a distorted bcc lattice ($cI16$ phase),
   surrounded by 
   a Fermi liquid, undergoes several structural transitions under high
   pressure, oscillating between a symmetric phase and a
   broken-symmetry dimerized phase. 
These phase oscillations seem to be a direct consequence of the non
   monotonic behaviour of the pair potential which is characterized by
   a Friedel wavelength $\pi/k_{\rm F}$ in the presence of a sharp Fermi
   sphere. 
Such findings could be relevant for the understanding of the tendency
   towards the formation of atomic pairs, which has been recently
   reported for the light alkali metals. 
We furthermore predict that a restoring of symmetry should take place
   at some stage under higher pressure, which would be signalled by a
   reentrant metallic character.

\begin{acknowledgement}
We thank N. W. Ashcroft, P. Ballone, N. H. March, G. Piccitto,
   P. S. Riseborough, K. Syassen for useful discussions and
   correspondence. 
\end{acknowledgement}

\bibliographystyle{mprsty}

\bibliography{a,b,c,d,e,f,g,h,i,j,k,l,m,n,o,p,q,r,s,t,u,v,w,x,y,z,zzproceedings,Angilella,notes}

\begin{thebibliography}{10}
\bibitem{note1}
Actually, this was a check that boundary effects were negligible, and that the
  lattice was large enough to reach full convergence.
\bibitem{note:Hafner}
See however Ref.~\protect\cite{Hafner:87} for a systematic derivation of
  response theory at all orders.
\end{thebibliography}

\end{document}